# Imaging through random media using coherent averaging

**Byungjae Hwang[1], Taeseong Woo[1], Cheolwoo Ahn[1], Jung-Hoon Park[1*]**

[1]Department of Biomedical Engineering, Ulsan National Institute of Science and Technology(UNIST), Ulsan 44919, Republic of Korea

Contacts:

Mr. Byungjae Hwang,

Email: paradoxx@unist.ac.kr

Dr. Taeseong Woo,

Email: taeseong@unist.ac.kr

Mr. Cheolwoo Ahn,

Email: carbonatedseason@unist.ac.kr

*Correspondence:

Dr. Jung-Hoon Park,

Email: jh.park@unist.ac.kr

Phone: +82-10-9240-3942

**Abstract**

We propose and demonstrate a new phase retrieval method for imaging through random media. Although methods to recover the Fourier amplitude through random distortions are well established, recovery of the Fourier phase has been a more difficult problem and is still a very active research area. Here, we show that by simply ensemble averaging shift-corrected images, the Fourier phase of an object obscured by random distortions can be accurately retrieved up to the diffraction limit. The method is simple, fast, does not have any optimization parameters, and does not require prior knowledge or assumptions about the sample. We demonstrate the feasibility and robustness of our method by realizing all computational diffraction-limited imaging through atmospheric turbulence as well as imaging through multiple scattering media.

**Introduction**

Imaging through random media is of paramount importance in many imaging scenarios where the acquired images are fundamentally distorted prior to acquisition, such as in imaging through atmospheric and oceanic turbulence, geophysical or biological media, and non-line-of-sight imaging [1-3]. To achieve this challenging feat, advances in hardware-based adaptive optics have shown great potential in recovering diffraction-limited resolution through dynamic disturbances [4-6]. However, hardware-based adaptive optics has strict requirements that must be fulfilled to guarantee optimal performance. Since the wavefront distortion must be known prior to correction, an effective guide star has to be available to measure the wavefront distortions and the total wavefront correction time has to fall safely within the decorrelation time of the random media. In addition, for efficient wavefront correction through highly scattering media such as thick biological tissue, the number of independent modes that must be corrected for satisfactory performance increases linearly with the signal level enhancement making fast wavefront correction technologically challenging [7-12].

To overcome such difficulties, computational image recovery using the information hidden in the distorted images have been demonstrated to be a powerful alternative in the field of astronomy [13-15]. Here, only the statistical properties of random distortions are used and there is no need to find and correct the exact wavefront distortions which greatly relaxes the complexity and constraints on the experimental system. Labeyrie first realized that short instants of turbulence do not erase the high spatial frequency information of an object but rather distorts this information [14]. By simply taking the ensemble average of the Fourier transform of the autocorrelation of multiple randomly distorted short exposure images, Labeyrie showed that we can retrieve the power spectrum of the original object as follows,

$$\left\langle \left| \tilde{I}_n(u,v) \right|^2 \right\rangle = \left| \tilde{O}(u,v) \right|^2 \cdot \left\langle \left| \tilde{P}_n(u,v) \right|^2 \right\rangle, \quad (1)$$

where $I_n$ is the $n$th acquired distorted image, $O$ is the object of interest and $P_n$ is the $n$th distorted point spread function (PSF) due to both the random distortion and intrinsic imaging system parameters, and tilde stands for the Fourier transform. In contrast, if we take the ensemble average of the Fourier transform of the short exposure images we obtain,

$$\left\langle \tilde{I}_n(u,v) \right\rangle = \tilde{O}(u,v) \cdot \left\langle \tilde{P}_n(u,v) \right\rangle. \quad (2)$$

Eq. (2) is just equivalent to taking the Fourier transform of a long exposure image where $\left\langle \tilde{P}_n(u,v) \right\rangle$ reaches zero for angles greater than $\frac{r_0}{\lambda}$, where $r_0$ is the Fried parameter. Labeyrie's key insight was that by simply taking the absolute square in the frequency domain prior to ensemble averaging, $\left\langle \left| \tilde{P}_n(u,v) \right|^2 \right\rangle$ is now non-zero up to the diffraction limit and enables recovery of the power spectrum of the object [14-17]. Due to the simplicity and robustness, Labeyrie's method has become a popular general method that is currently widely used to recover the Fourier power spectrum of an object through random disturbance.

However, recovery of the object requires not only the Fourier amplitude (square root of Fourier power spectrum) but also information about the Fourier phase. In this respect, variants of the Fienup type iterative phase retrieval methods have been widely used to recover the Fourier phase from only the Fourier power spectrum information [18-22]. The algorithm consists of iteratively imposing spatial domain constraints such as nonnegativity, known supporting regions of the object, or both, while using the recovered Fourier amplitude as the Fourier domain constraint. However, iterative phase retrieval algorithms need to solve a non-convex problem and are known to be very sensitive to almost all algorithm parameters such as prior assumed constraints, initial starting guesses and input-output scaling factors. In practice, many different initial guesses and/or input-output factors are often tested empirically until a satisfactory image is recovered (for example, see Figs. S1 and S2 in the

Supplementary Material). This brings limitations in computation speed and restricts establishing a single unique algorithm with fixed parameters that works robustly for all types of data generated in various experimental scenarios.

To surmount such problems, we can think about directly obtaining more information from the distorted images in addition to the Fourier power spectra. As these distorted images are still 'images', they naturally contain information about the Fourier phase as well. To obtain undistorted Fourier phase information, ensemble averaging the bispectrum of the distorted images has been shown to be an effective method [23-25]. The bispectrum is defined as the Fourier transform of the triple correlation. While the Fourier phase is lost in autocorrelation, triple correlation retains the phase information. The ensemble average of the bispectrum of the distorted images is expressed as,

$$\left\langle \tilde{I}_n^{(3)}(u_1,u_2,v_1,v_2) \right\rangle = \tilde{O}^{(3)}(u_1,u_2,v_1,v_2) \cdot \left\langle \tilde{P}_n^{(3)}(u_1,u_2,v_1,v_2) \right\rangle, \quad (3)$$

where the bispectrum of a function $G(x, y)$ is defined as,

$\tilde{G}^{(3)}(u_1,u_2,v_1,v_2) = \tilde{G}(u_1,v_1)\tilde{G}(u_2,v_2)\tilde{G}^*(u_1+u_2,v_1+v_2)$. When the coherent transfer function $C(u,v)$ of the random turbulence is a stationary random variable and the real and imaginary parts have zero-mean Gaussian distribution, $\left\langle \tilde{P}_n^{(3)}(u_1,u_2,v_1,v_2) \right\rangle$ is real and non-zero up to the diffraction limit and the Fourier phase of the object of interest can be obtained from the phase of the ensemble average of the bispectrum of distorted images [24]. In combination with the Fourier power spectrum retrieved from the autocorrelation of the distorted images in Eq. (1), the original object can be recovered. However, as we can see from Eq. (3), the bispectrum of a 2-D image is 4-D imposing high demands on computational memory. The object phase recovery from the bispectrum phase is also computationally expensive. To deal with such high computational loads, signal processing techniques using the properties of bispectrum [26-29], or accelerating computation through GPU based parallelization have been demonstrated [30-33]. However, widespread use of the bispectrum for phase retrieval is still largely limited due to the excessive computational load, slow speed, and the related difficulties in application to large images.

In this work, we propose a new framework for phase retrieval in imaging through random media. Similar to bispectrum analysis, the Fourier phase of an object is fully retrieved solely from the information contained in the randomly distorted images. Inspired by Labeyrie's method and the shift-and-add approach in astronomy [34-36], we demonstrate that ensemble averaging the shift-corrected distorted images (which we refer as coherent averaging of distorted images) can recover the missing Fourier phase information. The method is simple, fast, does not have any open parameters, is robust to noise, and does not fall into local minima. Our idea is grounded on the statistical properties of random waves where the autocorrelation of random turbulence is real and even-symmetric, which is a key property that is also crucial for bispectrum analysis.

**Results**

To describe the key principle for phase retrieval, we assume that the random coherent transfer function $C(\vec{k})$ induced by random media is statistically homogeneous (stationary) and isotropic, where $\vec{k}=(u,v)$. In other words, the ensemble average of the random coherent transfer function $<C(\vec{k})>$ is constant for all $\vec{k}$, and its autocorrelation $\Gamma_C(\vec{l})$ is only a function of the spatial frequency shift distance $\vec{l}$. Thus, the ensemble averaged optical transfer function ($OTF = \tilde{P}$), which is the autocorrelation of $C(\vec{k})$ can be written as,

$$OTF(\vec{l}) = \Gamma_C(\vec{l}) = <C(\vec{k})C^*(\vec{k}-\vec{l})>. \qquad (4)$$

Rewriting the coherent transfer function in its amplitude and phase, $C(\vec{k}) = A(\vec{k})e^{i\phi(\vec{k})}$, the $OTF$ can be expressed as,

$$OTF(\vec{l}) = <A(\vec{k})A(\vec{k}-\vec{l})e^{i\{\phi(\vec{k})-\phi(\vec{k}-\vec{l})\}}>. \qquad (5)$$

When the random magnitude and phase are mutually independent, the OTF can be further written as,

$$OTF(\vec{l}) = <A(\vec{k})A(\vec{k}-\vec{l})><e^{i\{\phi(\vec{k})-\phi(\vec{k}-\vec{l})\}}>. \qquad (6)$$

Since the phase distortion $\phi(\vec{k})$ is caused by a large number of random independent variables (random refractive index variations causing turbulence and multiple scattering), $\phi(\vec{k})$ and therefore $\phi(\vec{k})-\phi(\vec{k}-\vec{l})$ follow Gaussian statistics by the central limit theorem. The expectation value of the distorted phase term in Eq. (6) is then given as [37,38],

$$\left\langle e^{i\{\phi(k)-\phi(k-l)\}} \right\rangle = e^{-\frac{1}{2}\left\langle [\phi(k)-\phi(k-l)]^2 \right\rangle}. \qquad (7)$$

From Eqs. (6) and (7), we can see that the ensemble averaged OTF is a real and even function. Therefore, from Eq. (2), we see that the Fourier phase of the ensemble averaged randomly distorted images $\left\langle \tilde{I}_n(u,v) \right\rangle$ is equivalent to the Fourier phase of $\tilde{O}(u,v)$, the object we aim to recover. In other words, the reason we cannot recover the image of an object through turbulence using a long exposure image is not due to the scrambling of the Fourier phase, but rather the averaging out of the Fourier amplitude in Eq. (2) which erases the information. We can understand this intuitively by considering the fact that ensemble averaging multiple randomly distorted images results in a 'blurred' image where the high spatial frequency information is lost, rather than 'distorted', which would be caused by scrambling of the Fourier phase. Labeyrie conquered this problem by taking the absolute square of the Fourier domain information prior to averaging to save the Fourier amplitude. However, in this case, the Fourier phase information was lost during this process.

To overcome this limitation, let us consider the shift-and-add method [34,35]. In this process, we shift the images such that the distorted images, or equivalently, the random speckle PSFs for each imaged instance are aligned. Since the images are randomly distorted, all the individual realizations are caused by independent speckle PSFs. However, since the images are obtained with the same imaging system, the average speckle size is always decided by the full numerical aperture of the imaging system. Therefore, the speckle PSFs have a common correlation length given by the average speckle size. By aligning the random speckle PSFs in the shift-and-add process, the ensemble averaging of the shift-corrected speckle PSFs results in coherent averaging within the correlation length while areas beyond the correlation length are randomly averaged incoherently. In other words, the ensemble averaged shift-and-add speckle PSF consists of the sum of two parts; a sharp diffraction-limited focus and a diffuse background haze. Because of the background haze, ensemble averaged shift-and-add images have a common diffuse and blurry look which has been a severe drawback limiting widespread use of the shift-and-add method despite its simplicity. Here, we show that this limitation was due to the fact that the shift-and-add images were used as the final product in previous works. The hidden gem of this method, which has been somehow overlooked, lies in the Fourier spectra of the shift-and-add images. As we have previously mentioned, the Fourier phase of the ensemble averaged random OTF is zero which means that the Fourier phase of the object is not distorted. More importantly, although the shift-and-add OTF has a smaller Fourier amplitude for higher frequencies, it still does not fall to zero due to the sharp focus remaining above the background haze which allows us to directly retrieve the Fourier phase up to the diffraction limit.

To demonstrate the validity of our phase retrieval method, we acquired randomly distorted images in two different experimental schemes (Figure 1). In the first geometry, we imaged fine structured objects through atmospheric turbulence where random temporally independent distortions were applied. To create such an imaging condition, we imaged a U.S. five-dollar bill illuminated by a green LED with central wavelength of 565 nm through severe random temperature/density variations generated by a gas burner. In the second geometry, multiple random independent distortions were generated in the spatial domain using multiple scattering. A negative USAF target (Group number 1, Edmund Optics) was illuminated by a spatially incoherent narrowband light source which was generated by passing a coherent laser beam (532 nm, Shanghai Dream laser) through a speckle reducer (LSR-3005, Optotune). The scattered field then passed through a multiple scattering medium (ground glass, 220 Grit, Thorlabs) where the resulting distorted speckle image was recorded using a CMOS camera (LT545R, Lumenera) placed behind the scattering medium.

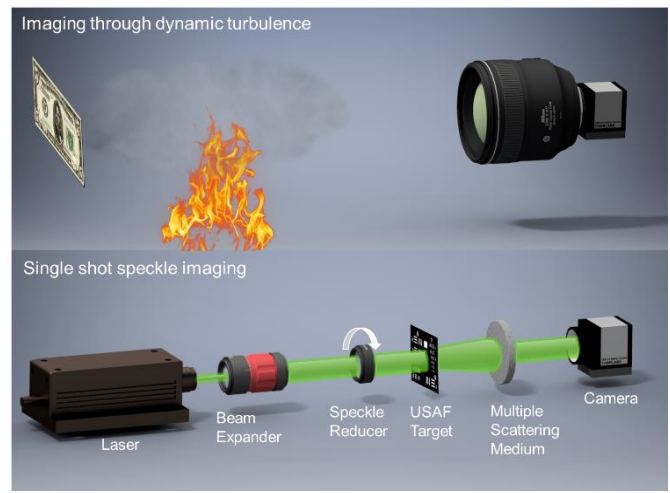

**Fig. 1 Experimental setup.** Two separate setups were used to demonstrate our method in (1) imaging through dynamic turbulence generated by fire and (2) single shot speckle imaging through multiple scattering.

The image recovery is first initiated by recovering the power spectrum of the object using the distorted images via Labeyrie's autocorrelation method [13,14]. We first checked that ergodicity is valid in our multiple scattering medium. Fig. 2a shows the original object and its autocorrelation. After passing through multiple scattering media, a fully developed speckle is imaged on the camera (Fig. 2b). By taking the autocorrelation of the speckle, the original object autocorrelation can be recovered due the optical memory effect (Fig. 2c) [39-43]. Furthermore, similar with holographic data storage, different subregions of the speckle pattern also hold information about the entire object, although with independent speckle footprints [44]. By dividing the full speckle pattern into subregions (60 x 60 pixels each), we performed autocorrelation of each subregion followed by ensemble averaging. As shown in Fig. 2d, the result is identical with that obtained using the full speckle as well as the original object Fourier spectra verifying that the ergodic nature of multiple scattering is valid in our experimental setup.

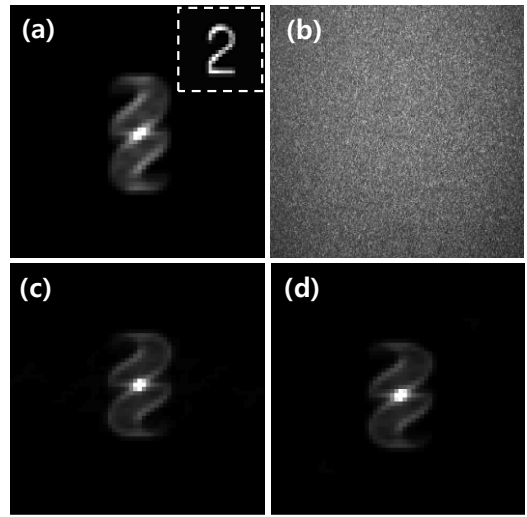

**Fig. 2 Recovery of object power spectra. a** The ground truth Fourier amplitude. The ground truth object is shown in the inset. **b** Imaged speckle pattern. **c** Autocorrelation of the speckle pattern **b**. **d** The speckle pattern **b** was divided into subimages with a size of 60 x 60 pixels each. The ensemble average of the autocorrelation of each subimages is identical with **a** and **c**.

The second half of image recovery performs the Fourier phase retrieval. Our key finding in this work is that simple shift-and-adding, or coherent averaging of the randomly distorted images actually holds information about the Fourier phase. To validate this property, we first imaged a sub-diffraction-limited pinhole as the test target. Short exposure images were acquired through severe dynamic turbulence using a gas burner to check its applicability in toughest conditions. Simply ensemble averaging the images resulted in a diffuse haze (Fig. 3a). The effective PSF for the long exposure is in the order of $\frac{\lambda}{r_0}$ and the spatial frequency cutoff is given by the inverse of the PSF. In stark contrast, coherent averaging results in a sharp focus on top of the same diffuse haze (Fig. 3b). The size of the sharp focus is indeed diffraction limited which proves that the spatial frequency cutoff is now extended up to the diffraction limit (the white dotted line in Fig. 3d is the diffraction-limited frequency cutoff as obtained when the turbulence was removed). As shown in Eqs. (6) and (7), the Fourier phase is constant for both ensemble averaged PSFs (Figs. 3c,d), however, with different cutoff frequencies (see Movie 1 and Fig S3. in the Supplemental Material for direct visualization of Fourier phase recovery upon coherent averaging and similar verification of resolution recovery for multiple scattering).

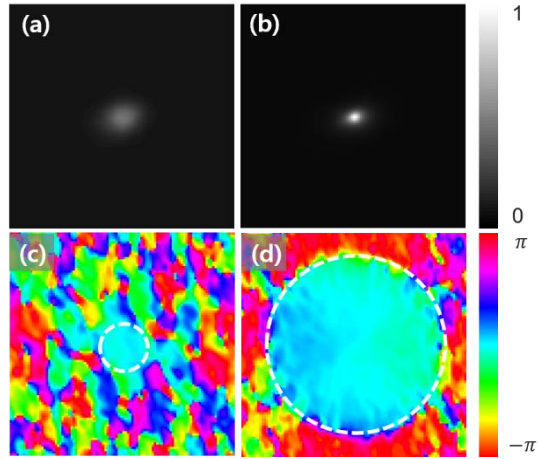

**Fig. 3 Comparison of simple averaging and coherent averaging. a** PSF obtained via simple averaging, **b** PSF via coherent averaging, **c** the Fourier phase of **a**, **d** the Fourier phase of **b**. The white dotted lines are guides to the eye showing that simple averaging is limited to a cutoff frequency defined by the Fried parameter. In contrast, the Fourier phase of coherent averaged PSF is constant up to the diffraction limit.

Based on our observation, we next proceeded to fully recover diffraction-limited images of extended objects. The entire workflow for the phase retrieval is illustrated in Fig. 4 as follows: (1) An arbitrary image is chosen as a reference image and cross correlation is performed between the reference image and all the other acquired images. (2) The relative shifts for all images are extracted and corrected for. (3) The shift-corrected images are ensemble averaged from which the Fourier phase is extracted. (4) The retrieved Fourier phase is combined with the Fourier amplitude obtained using Labeyrie's method and inverse Fourier transformed to obtain the recovered image. (5) Repeat steps (1-4) using the recovered image as the reference image until convergence (see Materials and Methods) or target iteration number.

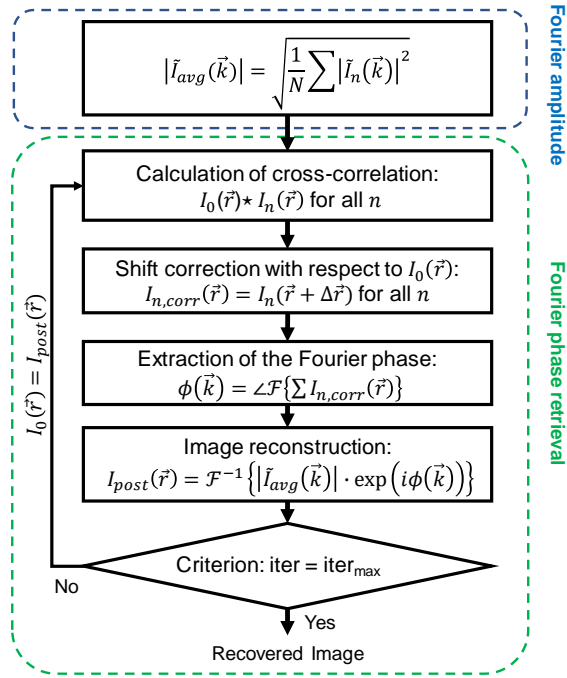

**Fig. 4 Reconstruction flowchart.** $\tilde{I}_n(\vec{k})$: Fourier transform of $I_n(\vec{r})$. $I_0(\vec{r})$: arbitrary image used as starting reference. $I_{n,corr}(\vec{r})$: shift-corrected $n$th image.

Figure 5 shows the final recovered images. All the images employed for the reconstruction were distorted by random turbulence (Fig. 5a). Averaging the acquired images is equivalent to a single long exposure image that just averages out the high spatial frequency information (Fig. 5b). By removing the relative shifts between the images and then averaging (conventional shift-and-add), we obtain an image that is slightly better than the simple sum (Fig. 5c). Although conventional shift-and-add method results in a degraded image due to incorrect Fourier amplitude recovery, the Fourier phase is fully recovered up to the diffraction limit. Using the Fourier phase from the shift-and-add image combined with the Fourier amplitude recovered using Labeyrie's method, we obtain the fully recovered image (Fig. 5d).

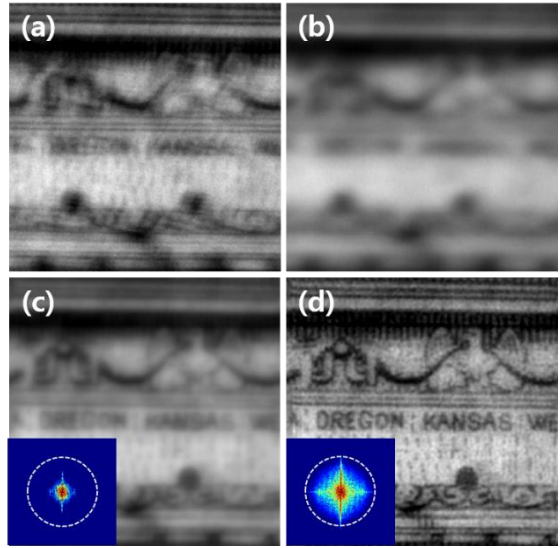

**Fig. 5 Imaging through turbulence. a** An example acquired image. **b** Simple ensemble averaging of acquired images. **c** Conventional shift-and-add ensemble averaging. **d** Our proposed method. Insets in **c** and **d** show the recovered Fourier amplitude compared with the diffraction limit outlined in yellow dotted lines.

Since our method relies on cross correlation between randomly distorted images, a natural question we can ask is whether this method can be applied robustly to severely distorted images where the cross correlation between images is expected to be minimal. To answer this question, we performed single shot experiments through multiple scattering media. As we anticipated, the acquired image was much more severely distorted compared to atmospheric turbulence. However, based on ergodicity of multiple scattering, a single image was sufficient to recover the object. Instead of using randomness realized as a function of time as in atmospheric turbulence, the randomness of multiple scattering in space was utilized by using subimages as independent realizations (Fig. 6a). When simply averaging all the subimages, information about the object was just averaged out resulting in an informationless constant background (Fig. 6b). After employing iterative coherent averaging of the distorted subimages (shift-and-add), useful information was retrieved similarly to imaging through turbulence (Fig. 6c). By using the Fourier phase from the iterative coherently averaged image in combination with the Fourier magnitude extracted from the ensemble averaging of distorted subimage autocorrelations, we successfully recovered the diffraction- limited resolution image (Fig. 6d) even through multiple scattering media.

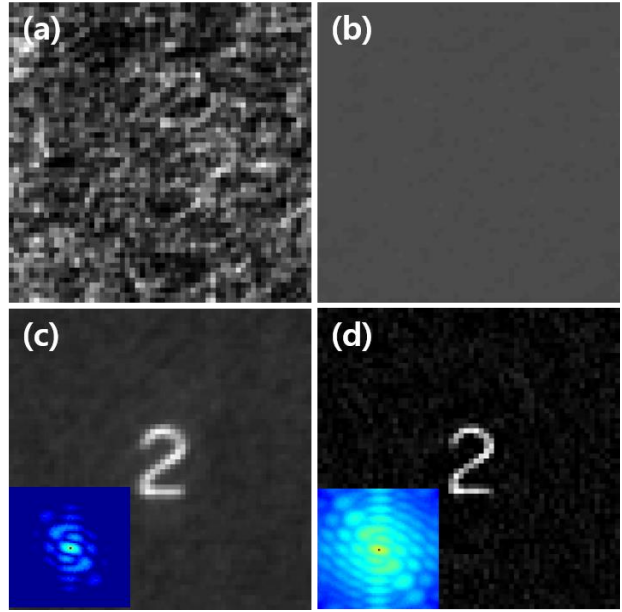

**Fig. 6 Imaging through multiple scattering media. a** An example acquired subimage. **b** Simple averaging of subimages. **c** Iterative coherent averaging of the acquired subimages. **d** Our proposed method. Insets in **c** and **d** show the recovered Fourier amplitude.

**Discussion**

We have demonstrated a new method for phase retrieval using randomly distorted images simply by exploiting the Fourier phase information that is recovered by using the shift-and-add method which has previously never been realized. To date, the biggest difficulty in realizing computational imaging through random media has been related to the accurate Fourier phase retrieval. In contrast to the well-established formalism for obtaining the Fourier amplitude which generally follows the initial idea proposed by Labeyrie, research related to developing effective iterative phase retrieval algorithms are still very actively ongoing. This reflects the difficulty of the problem and the fact that current existing techniques have various limitations that must be overcome. While previous approaches in iterative phase retrieval require assumptions about the object beforehand such as non-negativity and constraints on the extension area of the object, our method does not require preassumed constraints or initial guesses and is fast with almost no excess computational load. The process only consists of Fourier transforms and addition of the acquired images. It is also straightforward to parallelize the method for large images and makes this a powerful method for computational image reconstruction of extended field of views that are larger than the isoplanatic patch [45-47]. For example, a large object of interest where different areas have undergone different distortions can be independently analyzed and then simply recombined to obtain images that are larger than a single isoplanatic patch size. Using this approach, we demonstrate large field of view (940 x 730 pixels) real-time image reconstruction through severe turbulence generated by fire at an unprecedented refresh rate of 45 Hz (Movie 2). We have found the method to be robust in various degrees of randomness ranging from atmospheric turbulence to multiple scattering as well as related open datasets (see Fig. S4 in the Supplemental Material). The method has no open parameters that needs to be optimized and a single general-purpose algorithm (without any modifications) was found to be effective for various data sets. We believe that the simplicity, robustness, and accuracy of our method has potential to open new avenues for phase retrieval in randomly distorted data (see Fig. S5 and Table 1 in the Supplementary material for performance comparison with bispectrum analysis).

**Materials and methods**

Phase retrieval convergence

To demonstrate accurate phase retrieval, we first imaged an object through dynamic turbulence. As the turbulence was generated artificially using a gas burner, we could first turn off the turbulence to obtain a clean image of the object as the ground truth. Using the clean image, the Fourier phase difference between the ground truth and the recovered images through turbulence were obtained to check the accuracy of phase retrieval and also check that the phase retrieval is valid up to the diffraction limit of the imaging system. To validate the robustness of our method in the most challenging scenarios, we also made sure that the datasets used for reconstruction in imaging through turbulent media were always highly distorted by manually removing lucky guess images from the acquired datasets. Figure 7 depicts the convergence behavior of the retrieved Fourier phase for both imaging through atmospheric turbulence and through multiple scattering media, respectively. Using a single distorted image (Fig. 7a), the retrieved Fourier phase is inaccurate, especially for the high spatial frequency components. By increasing the number of shift-corrected randomly distorted images for ensemble averaging (Figs. 7b-d), the accurate Fourier phase is retrieved. For the level of turbulence generated by the gas burner, using less than 100 randomly distorted images and a single iteration was found to be enough for accurate phase retrieval (Movie 3). Next, phase retrieval using a test target in the shape of the number '2' was used for single shot phase retrieval through multiple scattering media. The ground truth phase to be recovered is shown in Figure 7e. For the single shot phase retrieval experiments, the total output speckle captured by the camera was divided into 133,533 overlapping subimages. Using a random subimage as the initial reference, we performed several iterations of our algorithm to recover the Fourier phase. Here, 10 iterations were found to be adequate to recover the correct phase (Figs. 7f-h, Movie 3).

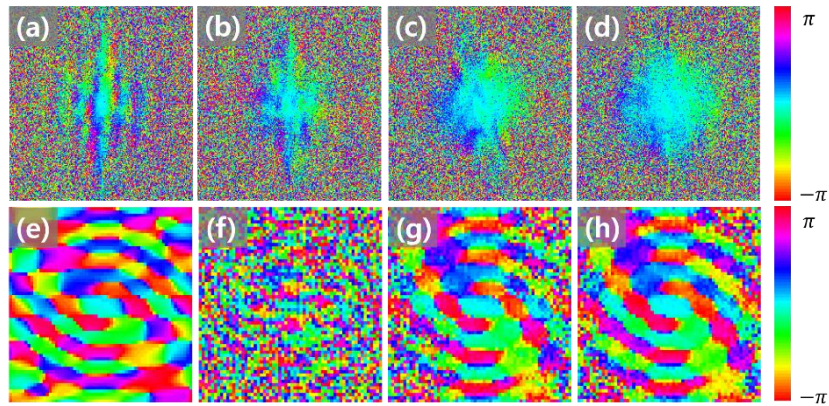

**Fig. 7 Fourier phase retrieval convergence.** Difference in Fourier phase between the ground truth and recovered images for imaging through atmospheric turbulence. **a** Using a single image. **b** 5 images. **c** 25 images. **d** 100 images. Imaging through multiple scattering media. **e** Ground truth Fourier phase. Recovered Fourier phase with, **f** 1 iteration, **g** 5 iterations, **h** 10 iterations.

**Acknowledgements**

This work was supported by the National Research Foundation of Korea (2017M3C7A1044966, 2019M3E5D2A01063812, 2020R1A6A3A01097999, 2021R1A2C3012903), and the TJ Park Foundation.

**Author contributions**

B.H. and J.P designed the project. B.H. developed the algorithm and performed experiments. W.T. and C.A. contributed analytical tools. B.H. and J.P. analyzed the data and wrote the manuscript with input from all authors. J.P. supervised the project.

**Conflict of interest**

The authors declare no competing financial interests.